\documentclass[]{PoS}
\usepackage{amsmath}

\def\eps{\varepsilon}

\newcommand{\be}{\begin{equation}}
\newcommand{\ee}{\end{equation}}
\newcommand{\bea}{\begin{eqnarray}}
\newcommand{\eea}{\end{eqnarray}}

\newcommand{\bi}{\begin{itemize}}
\newcommand{\ei}{\end{itemize}}

\newcommand{\vcb}{|V_{cb}|}
\newcommand{\vtd}{|V_{td}|}
\newcommand{\vub}{|V_{ub}|}
\newcommand{\vts}{|V_{ts}|}

\def\kpn{K^+\rightarrow\pi^+\nu\bar\nu}
\def\klpn{K_{L}\rightarrow\pi^0\nu\bar\nu}

\newcommand{\mev}{\, {\rm MeV}}

\title{CMFV models facing the recent progress in lattice calculations of
$B_{s,d}$ mixing}

\ShortTitle{CMFV models facing the recent progress in lattice calculations of
$B_{s,d}$ mixing}

\author{\speaker{Monika Blanke}\\
        {Institut fur Kernphysik, Karlsruhe Institute of Technology,
  Hermann-von-Helmholtz-Platz 1,
  D-76344 Eggenstein-Leopoldshafen, Germany}\\
 {Institut fur Theoretische Teilchenphysik,
  Karlsruhe Institute of Technology, Engesserstra\ss e 7,
  D-76128 Karlsruhe, Germany} \\
        E-mail: \email{monika.blanke@kit.edu}}

\abstract{Recent results by the Fermilab Lattice and MILC collaborations on the hadronic matrix elements entering $B_{s,d}-\bar B_{s,d}$ mixing have reached an unprecedented precision. Interestingly the Standard Model (SM) predictions using these updated values, together with the CKM elements obtained from tree-level decays, exhibit a significant tension with the measured values of the mass differences $\Delta M_s$, $\Delta M_d$ 
and their ratio. Assessing this tension in a model-independent way, it can be shown that models with Constrained Minimal Flavour Violation can not improve the situation with respect to the SM, in particular when the correlation with the CP-violating parameter $\varepsilon_K$ is taken into account. The new lattice results, if eventually confirmed by independent calculations, therefore imply the presence of new sources of flavour violation in the $\Delta F=2$ sector.}

\FullConference{9th International Workshop on the CKM Unitarity Triangle\\
		28  November -- 2 December 2016\\
		Tata Institute for Fundamental Research (TIFR), Mumbai, India}

\begin{document}


In the absence of direct signals of new particles, one of the most promising ways to discover beyond the Standard Model (BSM) physics is through its indirect contributions to low energy precision observables. Over the past years, flavour physics has become an important player in this field, due in particular to new experiments and an improved accuracy of the theoretical predictions. For the latter, the progress made in calculations of hadronic matrix elements on a discrete space-time lattice, has been crucial.

In early 2016, the Fermilab Lattice and MILC collaborations  presented new and improved results \cite{Bazavov:2016nty} for the hadronic matrix elements entering $B_{s,d}-\bar B_{s,d}$ mixings,
\begin{equation}\label{eq:FsqrtB}
 F_{B_s}\sqrt{\hat B_{B_s}}=(274.6\pm8.8)\mev,\qquad  F_{B_d} \sqrt{\hat B_{B_d}}=
(227.7\pm 9.8)\mev \,,
\end{equation}
as well as their ratio
\begin{equation}\label{eq:xi}
\xi=\frac{F_{B_s}\sqrt{\hat B_{B_s}}}{F_{B_d}\sqrt{\hat B_{B_d}}}=1.206\pm0.019\,.
\end{equation}
Determining $\Delta M_d$, $\Delta M_s$ and their ratio, using these values as well as the CKM matrix determined from tree-level decays as input, results in values that deviate by $1.8\sigma$, $1.1\sigma$ and $2.0\sigma$ from the data. Note that the tension is smaller when instead of the recent Fermilab-MILC results, the input values quoted by the Flavour Lattice Averaging Group \cite{Aoki:2016frl} are used. An independent confirmation of the results in \eqref{eq:FsqrtB} and \eqref{eq:xi} is therefore of utmost importance. 

Taking the observed tension at face value, it is instructive to analyse what kind of BSM physics is required to reconcile the $\Delta F=2$ sector with the data. To this end, its implications on the most minimal extension of the SM flavour sector, namely models with Constrained Minimal Flavour Violation (CMFV) \cite{Buras:2000dm,Buras:2003jf,Blanke:2006ig}, have been studied in \cite{Blanke:2016bhf}. In these models the SM Yukawa couplings are the only source of flavour and CP violation, and no new effective operators are generated beyond those that are already relevant in the SM. Consequently,  all CMFV contributions to the $\Delta F=2$ sector can be described by a single real and flavour-universal function $S(v)$ that replaces the SM loop function $S_0(x_t)$. It can be shown that
\cite{Blanke:2006yh}
\begin{equation}\label{eq:BBbound}
S(v)\ge S_0(x_t)= 2.322\,.
\end{equation}


\begin{figure}[b]
\centering{\includegraphics[width=.5\textwidth]{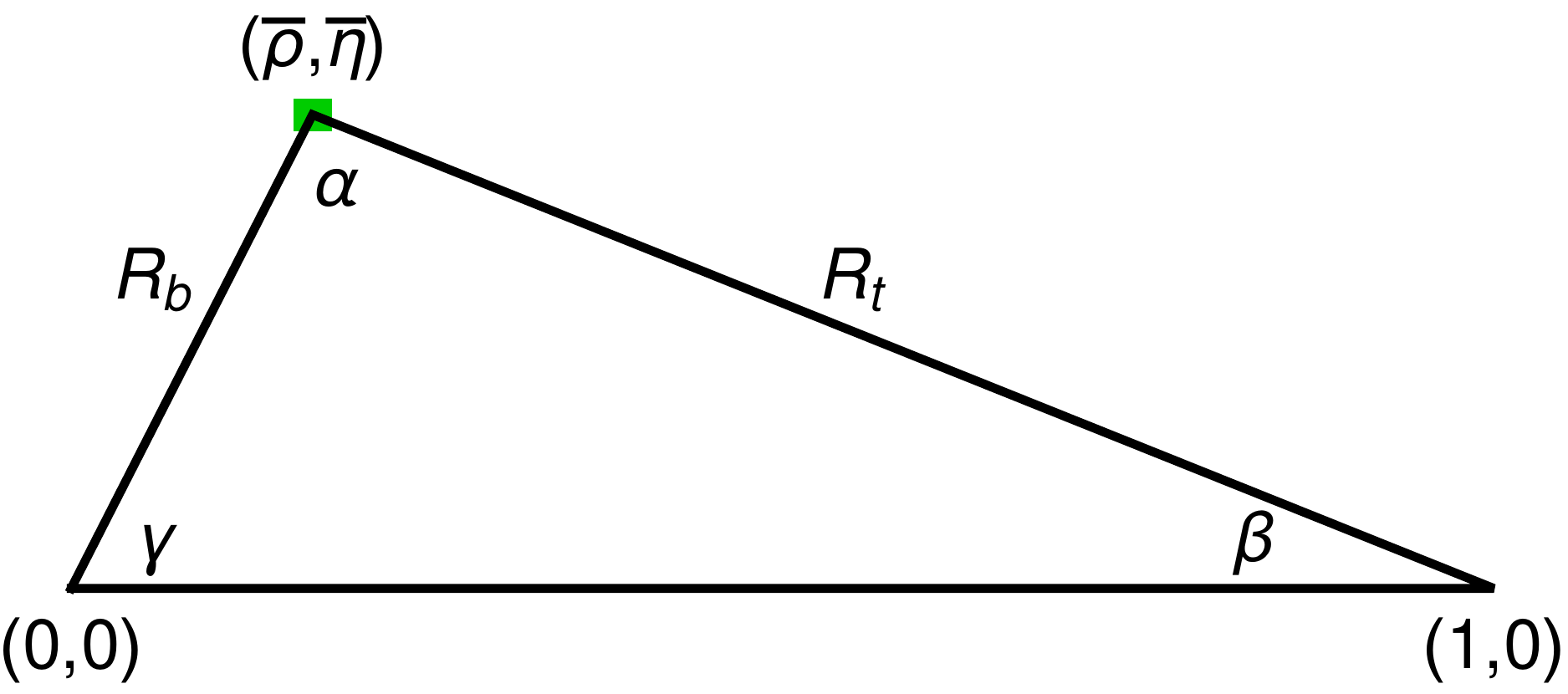}}
\caption{Universal unitarity triangle 2016. Figure taken from \cite{Blanke:2016bhf}.\label{fig:UUT}}
\end{figure}

In CMFV models the unitarity triangle can be constructed in a universal manner from $\Delta F=2$ observables \cite{Buras:2000dm}, using only the precisely known value of $|V_{us}|$ as tree-level input. The length of the side $R_t$ and the angle $\beta$ are determined by the ratio $\Delta M_d/\Delta M_s$ and the time-dependent CP asymmetry $S_{\psi K_S}$, respectively. Updating this determination using the most recent value of $\xi$ in \eqref{eq:xi}, one obtains an impressively precise picture for the universal unitarity triangle (UUT), as shown in Fig.~\ref{fig:UUT}, where the green rectangle displays the uncertainties in the apex of the UUT \cite{Blanke:2016bhf}.

\begin{figure}[t]
\includegraphics[width = 0.48\textwidth]{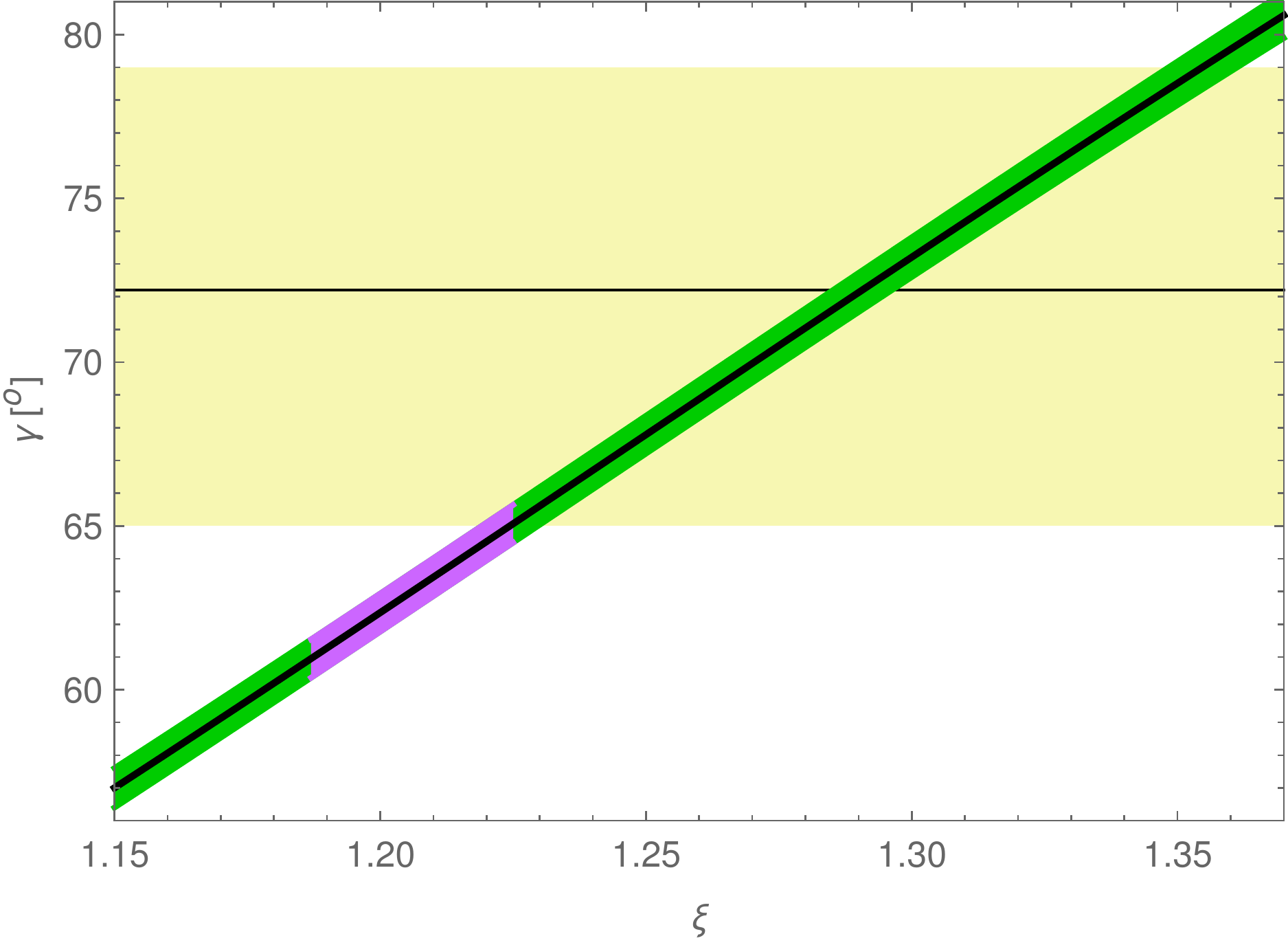}\hfill
\includegraphics[width = 0.48\textwidth]{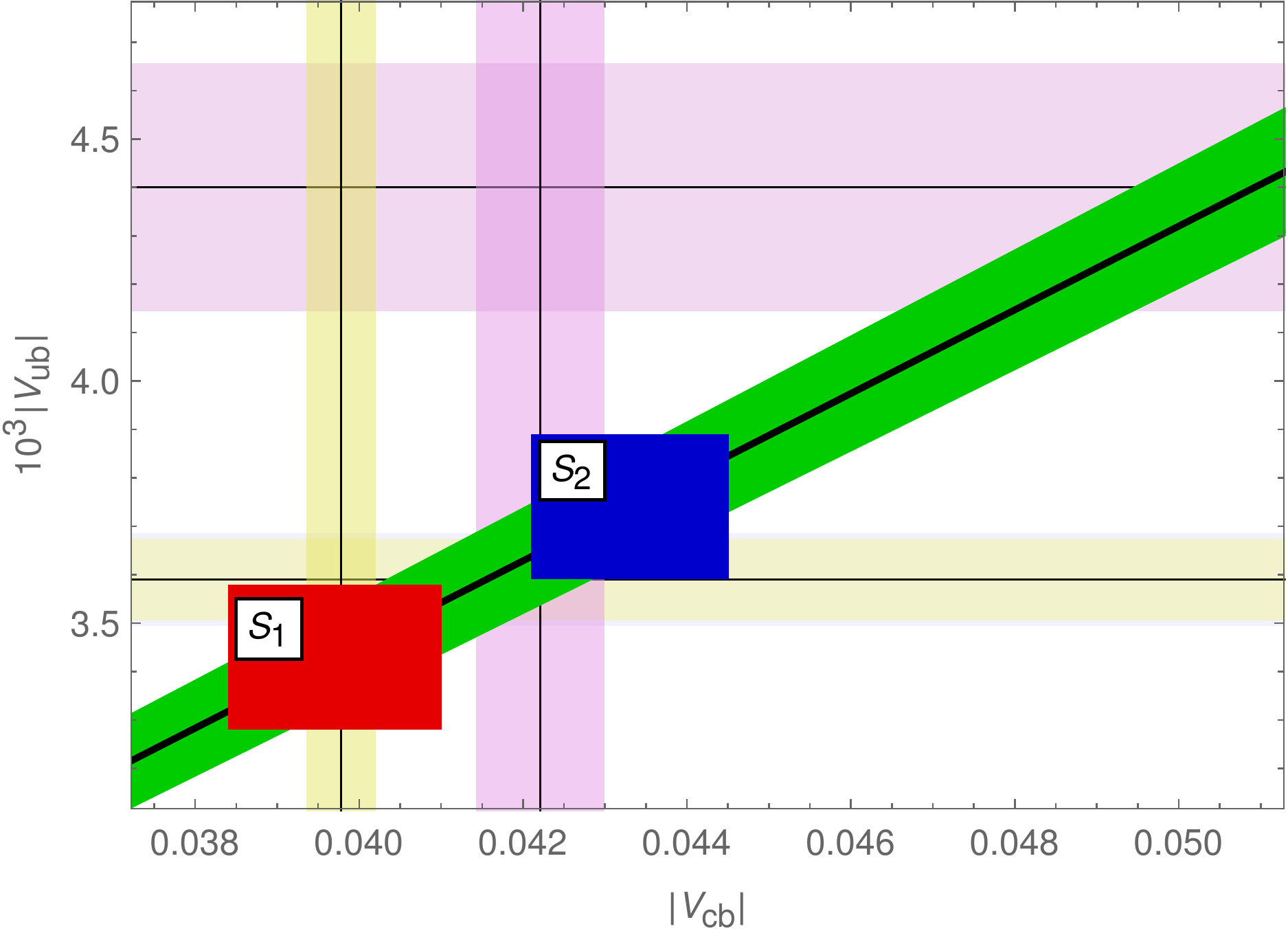}
\caption{{\it left:} CKM angle $\gamma$ as a function of the hadronic parameter $\xi$. The violet range corresponds to the new lattice determination of $\xi$ in \protect\eqref{eq:xi}, and the yellow range displays the tree-level determination of $\gamma$ in \protect\eqref{eq:gamma-tree}.
{\it right:} 
CKM elements $\vub$ versus $\vcb$ in CMFV (green) compared with the tree-level exclusive (yellow) and inclusive (violet) determinations. The squares display the results in strategies  $S_1$ (red) and $S_2$ (blue). Figures taken from \cite{Blanke:2016bhf}.
\label{fig:gamma-VubVcb}} 
\end{figure}

From the determined UUT one can deduce other CKM parameters, like the angle $\gamma$ or the ratio $|V_{ub}/V_{cb}|$. The results are shown in Fig.~\ref{fig:gamma-VubVcb} \cite{Blanke:2016bhf}. It is interesting to note that the obtained value for $\gamma$,
\begin{equation}\label{eq:gamma-UUT}
\gamma_\text{UUT}=(63.0\pm 2.1)^\circ\,,
\end{equation}
is below its tree-level value \cite{LHCb-gamma}
\begin{equation}\label{eq:gamma-tree}
\gamma_\text{tree} = (72.2^{+6.8}_{-7.2})^\circ\,.
\end{equation}
Moreover, the inclusive value of $\vub$ is clearly disfavoured in CMFV models. 

\begin{figure}[t]
\centering
\includegraphics[width = 0.48\textwidth]{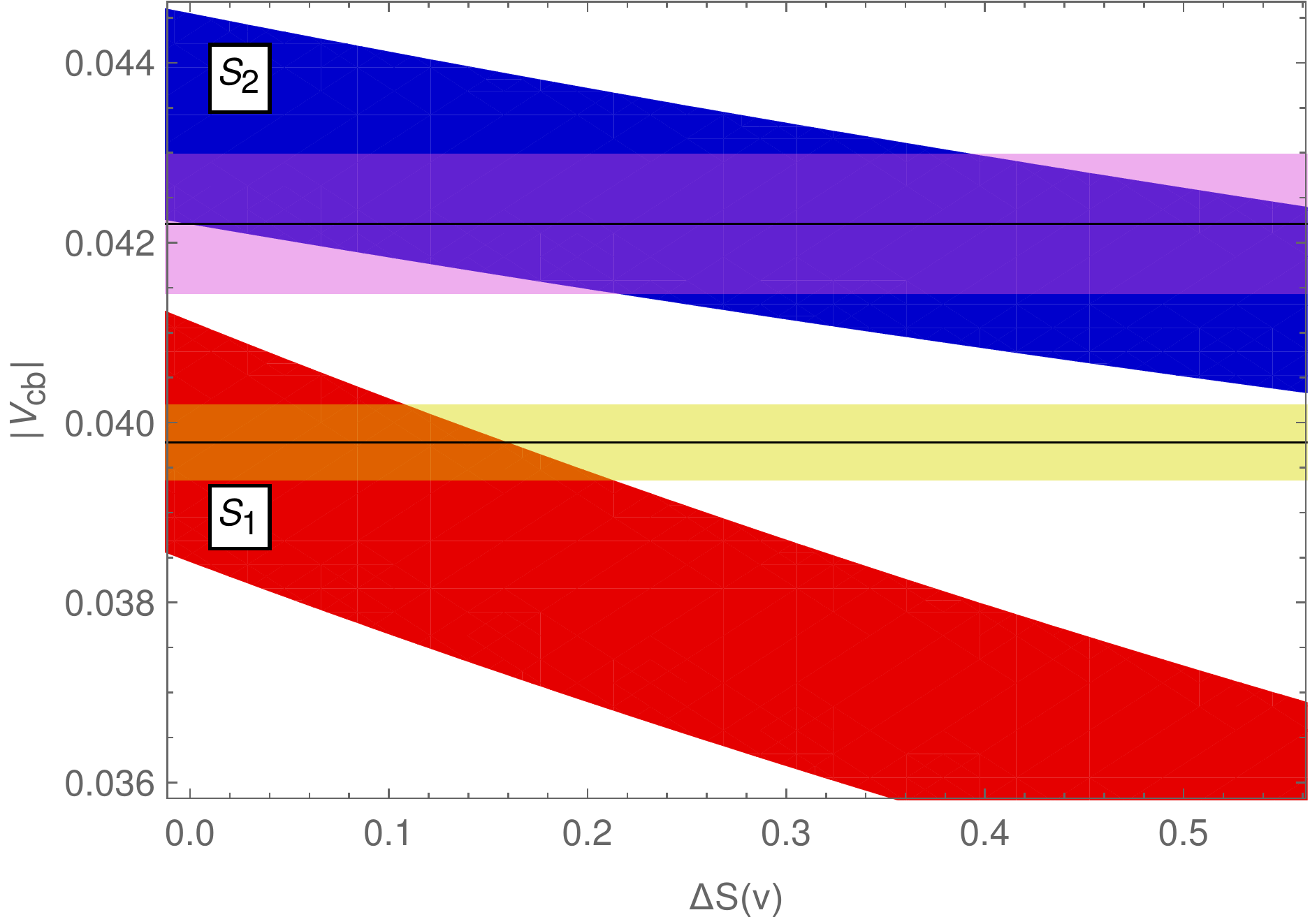}
 \caption{$\vcb$ versus {the flavour-universal BSM contribution} $\Delta S(v)$ obtained in $S_1$ (red) and $S_2$ (blue). The horizontal bands correspond to the inclusive (yellow) and exclusive (violet) tree-level measurements. Figure taken from \cite{Blanke:2016bhf}.\label{fig:Vcb-CMFV}}
\end{figure}

One more input is then required to determine $\vcb$ and thereby the full CKM matrix. As in all BSM models this is possible using tree-level charged current decays. However, due to the significant uncertainties and the tension between various determinations, we stay within the $\Delta F=2$ sector and determine $\vcb$ as a function of $S(v)$. This can be done using either $\Delta M_s$ (strategy $S_1$) or $\eps_K$ (strategy $S_2$) as input, with the outcome \cite{Blanke:2016bhf}
\begin{equation}
|V_{cb}|_{S_1} = (39.7 \pm 1.3)\cdot 10^{-3} \left[\frac{2.322}{S(v)}\right]^{1/2}\,,\qquad
|V_{cb}|_{S_2} =  (43.3 \pm 1.1)\cdot 10^{-3} \left[\frac{2.322}{S(v)}\right]^{1/4}\,.
\end{equation}
As can be seen from Fig.~\ref{fig:Vcb-CMFV}, these determinations yield inconsistent results, making apparent the tension between $\Delta M_{d,s}$ and $\eps_K$ present in CMFV models. It is interesting to note that, due to the lower bound in \eqref{eq:BBbound}, the tension is smallest in the SM limit $S(v) \equiv S_0(x_t)$ and can only be increased by new CMFV contributions. Even if the bound \eqref{eq:BBbound} was relaxed, which is in principle possible in some contrived CMFV scenarios \cite{Blanke:2006yh}, the tension would not disappear, but rather be shifted to a tension between the $\vcb$ values determined from $\Delta F=2$ processes and from tree-level decays. Flavour non-universal BSM contributions to the $\Delta F=2$ sector are therefore implied the Fermilab Lattice and MILC results in \eqref{eq:FsqrtB} and \eqref{eq:xi}.

Before having a closer look at BSM models beyond the CMFV hypothesis, it is instructive to consider the SM limits of the strategies $S_1$ and $S_2$. Table \ref{tab:CKM} collects the results for various CKM elements obtained using the strategies $S_1$ and $S_2$. As a consequence of the lower bound in \eqref{eq:BBbound}, these numbers serve as upper bounds in CMFV models. Table \ref{tab:rare} shows the SM predictions for various rare decay branching ratios, using the obtained CKM elements. Again we observe significant differences between the results of strategies $S_1$ and $S_2$ \cite{Blanke:2016bhf}.

\begin{table}[!h]
\centering
{
\begin{tabular}{|c|c|c|c|c|c|c|}
\hline
 & $\vts$ & $\vtd$ & $\vcb$ & $\vub$ &${\rm Im}\lambda_t$ & ${\rm Re}\lambda_t$        \\
\hline
$S_1$ &$39.0(13)$   & $8.00(29)$ & $39.7(13)$& $3.43(15)$ & $ 1.21(8)$& $-2.88(19)$\\
$S_2$ & $42.6(11)$  &$8.73(26) $ & $43.3(11) $& $3.74(14)$& $1.44(7)$ & $-3.42(18)$ \\
 \hline
\end{tabular}}
\caption{CKM elements in units of $10^{-3}$ and of $\lambda_t=V_{td} V_{ts}^*$ in units of $10^{-4}$ obtained using strategies 
$S_1$ and $S_2$ in the SM limit $S(v)=S_0(x_t)$. Table taken from \cite{Blanke:2016bhf}.
}\label{tab:CKM}
\end{table}
\begin{table}[!h]
\centering
{
\begin{tabular}{|c|c|c|c|c|}
\hline
  & $ {\mathcal{B}}(\kpn)$ & $ {\mathcal{B}}(\klpn)$ & $\overline{\mathcal{B}}(B_{s}\to\mu^+\mu^-)$ & $\mathcal{B}(B_{d}\to\mu^+\mu^-)$\\
\hline
$S_1$ &$7.00(71)\cdot 10^{-11}$   & $2.16(25)\cdot 10^{-11}$  &$3.23(24)\cdot 10^{-9}$ &$0.90(8)\cdot 10^{-10}$ \\
$S_2$ &$8.93(74)\cdot 10^{-11}$    &$3.06(30)\cdot 10^{-11}$   &$3.85(24)\cdot 10^{-9}$ & $1.08(8)\cdot 10^{-10}$\\
 \hline
\end{tabular}
}
\caption{SM predictions for rare decay branching ratios using the strategies 
$S_1$ and $S_2$. Table taken from \cite{Blanke:2016bhf}.
}\label{tab:rare}
\end{table}

As mentioned above, flavour non-universal BSM contributions are required to resolve the tension in the $\Delta F=2$ sector. In generic BSM models, the $\Delta F=2$ observables can be described by six independent new parameters. It is therefore always possible to fit the available data and bring the $\Delta F=2$ sector in agreement with the CKM elements measured in tree-level decays. Concrete models can then be tested by using the correlations between $\Delta F=2$ and $\Delta F=1$ processes which arise in specific scenarios, but are hidden in the effective theory picture. 

The more constrained models with a minimally broken $U(2)^3$ flavour symmetry \cite{Barbieri:2011ci,Barbieri:2012uh,Buras:2012sd,Blanke:2016bhf} can however be tested by means of $\Delta F =2$ observables and tree-level CKM elements only, without making use of $\Delta F=1$ rare decays. In this scenario, a triple correlation between the CP-asymmetry $S_{\psi K_S}$, the phase $\phi_s$ of $B_s-\bar B_s$ mixing and the ratio $|V_{ub}/V_{cb}|$ is predicted \cite{Buras:2012sd} and will be tested by future more accurate measurements. In addition, the low value of $\gamma$ in \eqref{eq:gamma-UUT}, that also holds in $U(2)^3$ models, may turn out to be problematic one day. 

In summary, the tension in the $\Delta F=2$ sector implied by the recent results by the Fermilab Lattice and MILC collaboration \cite{Bazavov:2016nty} has profound implications for the BSM flavour structure. As CMFV models have been shown \cite{Blanke:2016bhf} to always increase the tension, new sources of flavour and CP violation are required. Before being able to draw definite conclusions, we however have to wait for an independent cross-check of the lattice results for the hadronic matrix elements entering $B_{s,d}-\bar B_{s,d}$ meson mixing.

\vspace{3mm}

I am grateful to Andrzej J.\ Buras for the inspiring and pleasant collaboration which lead to the results presented in these proceedings.
My warm thanks go to the organisers of the CKM 2016 workshop in Mumbai, in particular the members of the local organising committee, who put a tremendous effort into making this workshop a successful and enjoyable one. 
My trip to CKM 2016 would not have been possible without the generous financial support by the DAAD Congress and Travel Programme of the German Academic Exchange Service.

\end{document}